\documentstyle[epsfig,aps]{revtex}
\widetext
\begin{document}
%\draft
\title{SUSY Search in TeV Scale Polarized Photon-Proton
Collisions}
\author{Z.Z.Aydin, A. Kandemir, O.Y\i lmaz and A.U. Y\i lmazer}
\address{Ankara University, Faculty of Sciences \\
Department of Engineering Physics \\ 06100 Tando\u{g}an, Ankara -
Turkey}

\maketitle
\begin{abstract}
Production of supersymmetric particles in TeV scale polarized
photon-proton collisions is discussed. Polarizations of both photon and
proton beams are considered. Associated productions of squark-chargino
and squark-gluino, and production of squark pairs have been examined.
Although the cross sections for different initial beam polarizations
do not differ much, the polarization asymmetry is sensitive to the
sparticle mass. We conclude that the capacity of future linac-ring
type TeV scale photon-proton colliders is quite promising in the search
for SUSY particles.
\end{abstract}

\section{Introduction}
\label{sec:intro}

   In recent years in addition to the well known TeV scale colliders
such as $pp(p\bar p),ep$ and $e^+e^-$ machines
the possibilities of the realization of $\gamma e$, $\gamma \gamma$
and $\gamma p$ colliders have been
proposed and discussed in detail \cite{milburn63}. Here one of the
main motivations is to reach the TeV scale at a subprocess level.
The collisions of protons
from a large hadron machine with electrons from a linac is the most
efficient way of achieving TeV scale at a constituent level in $ep$
collisions \cite{alekh87}. A further interesting feature is the possibility
of constructing $\gamma p$ colliders on the base of linac-ring
$ep$-machines. This can be realized by using the beam of high
energy photons produced by the Compton backscattering of laser photons
off a beam of linac electrons. Actually this method was originally
proposed to construct $\gamma e$ and $\gamma \gamma$ colliders on
the bases of $e^+e^-$ linacs
[Ginzburg et.al ref.1]. For the physics program at
$\gamma e$ and $\gamma \gamma$ machines see Refs.\cite{grifols84,goto90}.
Recently different
physics phenomena which can be investigated at $\gamma p$ colliders
have been considered in a number of papers \cite{alekh91,buch93,atag94}.
It seems that these
machines may open new possibilities for the investigations of the
Standard Model and beyond it. For a review see Ref.\cite{zaetal96}.

On the other hand, among the various extensions beyond
the SM the supersymmetry idea seems to be a well-motivated strong
candidate to investigate the new physics around TeV scale
\cite{haber86}. In SUSY
inspired models the usual particle spectrum is doubled at least;
every particle has a superpartner differing in spin by a half unit.
Also two Higss doublets are needed to give mass to both up and down
quarks and make the theory anomaly free.
It is highly believed that superpartners of the known particles
should have masses below 1 TeV in order not to loose the good
features of the SUSY. The direct experimental
evidence for the sparticles is still lacking and the results of
the experiments in the existing colliders indicate that squarks
have masses $m_{\tilde q}\geq 130 GeV$; consequently higher energy
scale should be probed and it is desirable to reach the TeV scale at
a constituent level
\cite{harv92,baer95}. Therefore it might be sensible to say that HERA,
LEP, FERMILAB and LHC should be sufficient to check the idea of
low energy SUSY, namely the scale between 100 GeV and
1 TeV, however experiments
at all possible types of colliding beams would be inevitable to
explore the new physics at the TeV scale.

   In this report we will study the productions of SUSY particles
in polarized TeV scale photon-proton collisions.
Several SUSY production processes such as
$\gamma p\rightarrow\tilde{q}\tilde{w}X$,
$\gamma p\rightarrow\tilde{w}\tilde{w}X$,
$\gamma p\rightarrow\tilde{q}\tilde{g}X$,
$\gamma p\rightarrow\tilde{q}\tilde{\gamma}X$
(or $\tilde{q}\tilde{z}$) and
$\gamma p\rightarrow\tilde{q}\tilde{q}^*X$
have already been discussed \cite{buch93} without taking into account
the initial beam polarizations. Polarization effects have been
investigated in \cite{pol96}.
Also scalar leptoquark
productions at TeV energy $\gamma p$ colliders have been investigated
\cite{atag94}. Let us first briefly review the polarization
properties of high energy photon beams.
\bigskip

\section{Polarized High Energy Gamma Beams}

A beam of laser photons ($\omega_{0}\approx 1.26$ eV, for example) with high
intensity, about $10^{20}$ photons per pulse, is Compton-backscattered off
high energy electrons ($E_e$=250 GeV, for example) from a linear accelerator
and turns into hard photons with a conversion coefficient close to unity.
The energy of the backscattered photons, $E_{\gamma}$, is restricted by the
kinematic condition $y_{max}= 0.83$ (where $y=E_{\gamma}/E_{e}$) in order
to get rid of background effects, in particular $e^{+}e^{-}$ pair production
in the collision of a laser photon with a backscattered photon in the
conversion region.

The details of the Compton kinematics and calculations of the cross section can
be found in ref.\cite{milburn63}
(see Ginzburg et.al). The energy spectrum of the high energy
real (backscattered) photons, $f_{{\gamma}/e}(y)$, is given by

\begin{eqnarray}
f_{{\gamma}/e}(y)=\frac{1}{D(\kappa)}\biggr[1-y+\frac{1}{1-y}
-4r(1-r)-\lambda_e\lambda_0 r\kappa(2r-1)(2-y)\biggr]
\end{eqnarray}
where $\kappa=4E_e\omega_0/m_e^2$ and $r=y/\kappa(1-y)$.
Here $\lambda_0$ and $\lambda_e$ are the laser photon and
the electron helicities respectively, and $D(\kappa)$ is

\begin{eqnarray}
D(\kappa)&=&\biggr(1-\frac{4}{\kappa}-\frac{8}{\kappa^2}\biggr)
ln(1+\kappa)+\frac{1}{2}+\frac{8}{\kappa}-\frac{1}{2(1+\kappa)^2}\nonumber\\
&+&\lambda_e\lambda_0\biggr[(1+\frac{2}{\kappa})ln(1+\kappa)-
\frac{5}{2}+\frac{1}{1+\kappa}-\frac{1}{2(1+\kappa)^2}\biggr]
\end{eqnarray}
In our numerical calculations, we assume $E_e\omega_0=0.3$ $MeV^2$ or
equivalently $\kappa=4.8$ which corresponds to the optimum value of $
y_{max}=0.83$, as mentioned above.

The energy spectrum, $f_{\gamma /e}(y)$, does essentially depend on the value
$\lambda _e\lambda _0$. In the case of opposite helicities ($\lambda
_e\lambda _0=-1$) the spectrum has a very sharp peak at the high
energy part of the photons.
This allows us to get a highly monochromatic high energy gamma
beam by eliminating low energy part of the spectrum \cite{milburn63}
(see Ginzburg et.al and  Borden et.al).
On the contrary, for the same helicities ($\lambda _e\lambda
_0=+1 $) the spectrum is flat.

The average degree of linear polarization of the photon is proportional to
the degree of linear polarization of the laser. In our calculations, we
assume that the degree of linear polarization of the laser is zero so that
the final photons have only the degree of circular polarization $(\lambda
(y)=<\xi _2>\neq 0$ and $<\xi _1>$$=<\xi _3>=0$ $)$. The circular
polarization of the backscattered photon is given as follows

\begin{eqnarray}
<\xi _2>=\lambda(y)=\frac {(1-2r)(\frac{1}{1-y}+1-y)\lambda_0
+\lambda_e r\kappa \biggr(1+(1-2r)^2(1-y)\biggr)}{\frac{1}{1-y}+1-y-4r(1-r)
-\lambda_0\lambda_e r\kappa(2r-1)(2-y)}
\end{eqnarray}

For the same initial polarizations ($\lambda_0\lambda_e=+1$) , it is seen
that $\lambda(y)\approx +1$, as nearly independent of $y$;
while for the case of the opposite
polarizations ($\lambda_0\lambda_e=-1$),
the curve $\lambda(y)$
smoothly changes from $-1$ to $+1$ as $y$ increases from zero to $0.83$
 \cite{milburn63}.

\section{Pair Production of Squarks}
\label{sec:pro}
    We consider the minimal supersymmetric extension of the standard
model (MSSM) which includes soft breaking terms. Here we have only
taken into account the direct interaction of the photons with partons
in the proton. The resolved contributions to the SUSY particle
productions are relatively small, around 10\% which do not affect
our estimates for the discovery mass limits of new particles very much
\cite{drees85}.

    The invariant amplitude for the subprocess
$\gamma g\rightarrow\tilde{q}\tilde{q}^*$ proceeds via the
t-, u-channel squark exchange
and four-point vertex interactions. The polarized
differential cross-section of this
subprocess can be calculated in terms of the usual Mandelstam variables
$\hat s=(p_1+p_2)^2, \hat t=(p_1-p_3)^2$ and $\hat u=(p_1-p_4)^2 $ as  :

\begin{eqnarray}
\frac{d\hat{\sigma}}{d\hat{t}}=
\biggr(\frac{d\hat{\sigma}}{d\hat{t}}\biggr)_{np}
+\xi_2 \eta_2\biggr(\frac{d\hat{\sigma}}{d\hat{t}}\biggr)_{pol}
\end{eqnarray}

\begin{eqnarray}
\biggr(\frac{d\hat{\sigma}}{d\hat{t}}\biggr)_{np}=\frac{e^2e^2_{\tilde{q}}
g^2_{s}}{8\pi\hat{s}^2}\biggr[&&1+
\frac{m^{2}_{\tilde{q}}
(3m^{2}_{\tilde{q}}-\hat{s}-\hat{t})}{2(\hat{s}+
\hat{t}-m^2_{\tilde{q}})^2}+\frac{(3m^2_{\tilde{q}}-
\hat{s}+\hat{t})}{4(m^2_{\tilde{q}}-\hat{t})}\nonumber \\
&&+\frac{m^2_{\tilde{q}}(m^2_{\tilde{q}}+\hat{t})}
{(m^2_{\tilde{q}}-\hat{t})^2}+
\frac{(5m^2_{\tilde{q}}+2\hat{s}-\hat{t})}
{4(\hat{s}+\hat{t}-m^2_{\tilde{q}})}+\frac{(\hat{s}-2m^2_{\tilde{q}})
(\hat{s}-4m^2_{\tilde{q}})}{4(m^2_{\tilde{q}}-\hat{t})(\hat{s}+
\hat{t}-m^2_{\tilde{q}})}\biggr]
\end{eqnarray}

\begin{eqnarray}
\biggr(\frac{d\hat{\sigma}}{d\hat{t}}\biggr)_{pol}=\frac{e^2e^2_{\tilde{q}}
g^2_{s}}{16\pi\hat{s}^2}\biggr[\frac{m^2_{\tilde{q}}(\hat{s}-2\hat{t})
+m^4_{\tilde q}+\hat{s}\hat{t}+{\hat{t}}^2}{(m^2_{\tilde q}-\hat{s}
-\hat{t})(m^2_{\tilde q}-\hat{t})}\biggr]
\end{eqnarray}
Here $e_{\tilde q}$ is the squark charge, $g_s=\sqrt{4\pi \alpha_s}$
is the strong coupling
constant, $e=\sqrt{4\pi \alpha}$, $\alpha$ is the fine structure
constant and $\xi_2, \eta_2$ are the circular polarization parameters
of the photon and gluon respectively.
After performing the integration over $\hat{t}$ one can easily
obtain the total cross section for the subprocess
$\gamma g\rightarrow\tilde{q}\tilde{q}^*$ as follows :

\begin{eqnarray}
\hat{\sigma}(\hat{s},&m_{\tilde{q}}&,\xi_2,\eta_2)=\hat{\sigma}_{np}
+\xi_2\eta_2\hat{\sigma}_{pol}\\
\hat{\sigma}_{np}&=&\frac{\pi\alpha e^2_{\tilde{q}}
\alpha_s}{2\hat{s}}\biggr[2\beta(2-\beta^2)-(1-\beta^4)
ln\frac{1+\beta}{1-\beta}\biggr]\\
\hat{\sigma}_{pol}&=&\frac{\pi\alpha e^2_{\tilde{q}}
\alpha_s}{2\hat{s}}\biggr[2\beta-2(1-\beta^2)
ln\frac{1+\beta}{1-\beta}\biggr]
\end{eqnarray}
where $\beta=(1-4m^2_{\tilde{q}}/\hat{s})^{1/2}$ .
The above expression in (4) is a cross section for left- squarks
or right-squarks only, and a color factor of $C=1/2$ is already
included. In order to obtain the total cross section for the process
$\gamma p\rightarrow\tilde{q}\tilde{q}^*X$ one should integrate
$\hat{\sigma}$ over the gluon and photon distributions. For this
purpose we make the following change of variables: first
expressing $\hat{s}$ as $\hat{s}=x_1x_2s$ where
$\hat{s}=s_{{\gamma}g}$, $s=s_{ep}$, $ x_1=E_{\gamma}/E_e$,
$x_2=E_g/E_p$ and furthermore calling $\tau =x_1x_2$, $x_2=x$
then one obtains $dx_1dx_2 = dx d\tau/x$. The limiting values
are $x_{1,max}=0.83$ in order to get rid of the background effects
in the Compton backscattering, particularly $e^+e^-$ pair
production in the collision of the laser with the high energy
photon in the conversion region,
$x_{1,min}=0$, $x_{2,max}=1$,
$x_{2,min}=\frac{\tau}{0.83}$, $\hat{s}_{min}=4m_{\tilde{q}}^2$.
Then we can write the total cross section with
right circular polarized laser and spin-up proton beam
polarized longitudinally as follows:

\begin{eqnarray}
\sigma_{R\uparrow}=\int^{0.83}_{4m^2_{\tilde{q}}/s}
&d\tau&\int^{1}_{\tau/0.83}
dx\frac{1}{x}
\biggr\{
P\biggr[f^{\gamma}_{+}(\frac{\tau}{x})f^g_{+/\uparrow}(x)
\hat{\sigma}
(\tau s,m_{\tilde{q}},1,-1)
+f^{\gamma}_{+}(\frac{\tau}{x})f^g_{-/\uparrow}(x)
\hat{\sigma}(\tau s,m_{\tilde{q}},
1,1)\biggr]\nonumber\\
&+&(1-P)\biggr[f^{\gamma}_{+}(\frac{\tau}{x})f^g_{+/\downarrow}(x)
\hat{\sigma}
(\tau s,m_{\tilde{q}},1,1)
+f^{\gamma}_{+}(\frac{\tau}{x})
f^g_{-/\downarrow}(x)\hat{\sigma}(\tau s,m_{\tilde{q}},
1,-1)\biggr]\biggr\}
\end{eqnarray}
where the $(\pm 1,\pm 1)$ inside $\hat{\sigma}$ refers to the values of
the Stokes parameters $(\xi_2,\eta_2)$ respectively.
Using the Eq.(4) the above expression yields
%\newpage
\begin{eqnarray}
\sigma_{R\uparrow}=\int^{0.83}_{4m^2_{\tilde{q}}/s}
d\tau\int^{1}_{\tau/0.83}
dx\frac{1}{x}\biggr\{
P\biggr[f^{\gamma}_{+}(\frac{\tau}{x})f^g_{unpol}(x)
\hat{\sigma}_{np}
+f^{\gamma}_{+}(\frac{\tau}{x})(-\Delta{f}^g_{pol}(x))
\hat{\sigma}_{pol}\biggr]\nonumber\\
+(1-P)\biggr[f^{\gamma}_{+}(\frac{\tau}{x})f^g_{unpol}(x)
\hat{\sigma}_{np}
+f^{\gamma}_{+}(\frac{\tau}{x})(+\Delta{f}^g_{pol}(x))
\hat{\sigma}_{pol}\biggr]\biggr\}
\end{eqnarray}
where P is the polarization percentage of spin-up protons in
the beam and for a maximum attainable degree of 70\% longitudinal
polarization
$P=0.85$. Also $f^{\gamma}_{\pm}$ and
$f^g_{\pm}$ are the polarized distributions for the
photon and gluon inside the proton, respectively.
 In the numerical calculations
we used the unpolarized and difference distributions,
$f^g_{unpol}$ and $\Delta{f}^g_{pol}$ taken from
\cite{eich84}:

\begin{eqnarray}
f^g_{unpol}(x)=\frac{1}{x}\biggr[(2.62+9.17x)(1-x)^{5.90}\biggr]
\end{eqnarray}
and
\begin{eqnarray}
\Delta{f}^g_{pol}=16.3001x^{-0.3}(1-x)^7
\end{eqnarray}
The energy spectrum of the high energy real photons
$f_{\gamma}(y)$ is given as in Eq.(1-2). To carry out
the numerical integration we take $e_{\tilde q}=2/3$
hence consider only u-type squarks and $\alpha_s=0.1$ is taken.
The results of the numerical calculations clearly indicate that
the process $\gamma p\rightarrow\tilde{q}
{\tilde{q}}^*X$ has a detectable cross section. In
Figs.~\ref{fig1}(a-c)
the dependence on the squark masses is shown for different
$\gamma p$ colliders. The relevant parameters of these machines
are given in Table 1.

On the other hand observation limit for the new particles
is taken to be 100 events per running year  ($10^7$ seconds).
This level of observability is considered to be satisfactory since
the background is expected to be clearer than that encountered
in the hadron colliders where 1000 events/year is usually desired
due to the strong background processes. Hence taking into account
the luminosity values given in Table~\ref{tab1}
[see Aydin et al. in ref.1],
one can easily find from the Figs.~\ref{fig1}(a-c) the upper mass
limits for the squarks. These values of discovery limits are also
tabulated in Table~\ref{tab1}. For the d-type squarks
$(\tilde{d}_{L,R}, \tilde{s}_{L,R}, \tilde{b}_{L,R})$
putting $e_{\tilde{q}}=-1/3$ , results in the decrease of the
cross section by a factor 1/4 , but the corresponding upper mass
values are approximately some seventy percent of those in the last
column of the Table~\ref{tab1}. Furthermore the mixings among left
and right squarks might be taken into consideration. Because their
masses are expected to be close to each other, one can assume
that $\tilde{q}_L$ and $\tilde{q}_R$ are degenerate in mass
which points out that for the $\tilde{q}{\tilde{q}}^*$
production one can always consider the sum of the two cross
section for $\tilde{q}_L$ and $\tilde{q}_R$ production.
Hence if we do this incoherent sum of $\tilde{q}_L$ and $\tilde{q}_R$
and with $m_{{\tilde{q}}_L}=m_{{\tilde{q}}_R}$ we must
multiply the expression in (4) by an extra factor two,
which increases the discovery limits for the
squark masses. One additional  assumption can be made by assuming
five degenerate flavours of $\tilde{q}_L$ and $\tilde{q}_R$,
(namely except stop) which multiplies the individual
squark pair production
cross section, Eq.(3) by about a factor of six and a half. We
see that the cross sections for different initial beam polarizations
do not differ much which reflects the scalar nature of the squarks.

On the other hand the use of the polarized beams make it possible
to look for the polarization asymmetries as a function
of the squark masses. In Figs.~\ref{fig2}(a-c) we have presented
the results of the polarization asymmetry using the results of the
up and down polarized proton beams. Asymmetry has been defined through
the following relation

\begin{eqnarray}
A_{\uparrow \downarrow}=\frac{\sigma_{R\uparrow}-\sigma_{R\downarrow}}
{\sigma_{R\uparrow}+\sigma_{R\downarrow}}
\end{eqnarray}
As can be seen from the Figures 2(a-c) the asymmetry is
sensitive to the
squark mass which can be useful for determination of the
mass parameter.
Furthermore another asymmetry might be defined considering the
opposite polarizations of the laser beam. Hence left-right
asymmetry defined with respect to the laser beam as

\begin{eqnarray}
A_{LR}=\frac{\sigma_{R\uparrow}-\sigma_{L\uparrow}}
{\sigma_{R\uparrow}+\sigma_{L\uparrow}}
\end{eqnarray}
The results of the calculations are plotted in Fig.~\ref{fig3} and
a similar behaviour is seen.\\
{\bf {\it Signature for the pair production of squarks:}} \rm One
characteristic feature of the R-parity conserving
supersymmetric processes
is the large missing energy. Usually photino mass is assumed to
be less than $m_{\tilde{q}}$. This immediately implies that
$\tilde{q}\rightarrow q\tilde{\gamma}$. The decay of the
squark into photino is not the only possibility. For squark
masses larger than 200-300 GeV and depending on the mass
spectrum of the other SUSY particles there exist additional
decay modes such as $\tilde{q}\rightarrow q\tilde{g}$ ,
$\tilde{q}\rightarrow q\tilde{w}$ ,
$\tilde{q}\rightarrow \tilde{q}W $,$\tilde{q}\rightarrow q\tilde{z}$.
Branching ratios of these decays depend on the masses of
these particles and coupling constants.
One possibility of the decay of zino is the decay into a
neutrino and a sneutrino. Both particles will not be observed like
photinos. Therefore  the signature for the process
$\gamma p\rightarrow\tilde{q}{\tilde{q}}^*X$
will be in general multijets + lepton(s) + large missing energy and
missing $p_T$. The definite polarization asymmetries associated
with the missing energy and momentum may help in separating these
events from the backgrounds.

Actually the production of the squark pairs
in $ep\rightarrow{\gamma}^* g
\rightarrow\tilde{q}\tilde{q}^*$ collisions via the quasi-real photon
gluon fusion were studied before
\cite{drees85,gae90}. However in these studies Weizsacker-Williams
approximation has been used for the quasi-real photon distribution
$f_{\gamma/e}$. Since the
WW-spectrum is much softer than the real $\gamma$ spectrum
the discovery mass limits for the squarks in our case turn out
to be much higher than the conventional ep-colliders.

   Finally our analysis shows that the future $\gamma p$ colliders
can have considerable capacities in addition to the well known $pp$ and
$e^+e^-$ colliders in the investigation of supersymmetric particles.
We see that the range of squark masses that can be explored at
various $\gamma p$-machines (200 GeV - 0.85 TeV) are higher than
the corresponding values at standard type
$ep$-colliders (20 GeV - 80 GeV for HERA).
Several LHC studies have shown that the reach for squarks will be
greater than 1 TeV. But clearer backgrouds in a
gamma-proton collider can be considered to be an advantageous
feature in extracting supersymmetric signals.

\section{Gluino-Squark Production}

In this section, the subprocess
$\gamma\,q\longrightarrow \tilde{q}\,\tilde{g}$ is considered taking into
account the direct interaction of the photons with partons in the proton.
The invariant amplitude for this subprocess is the sum of the
terms corresponding to the s-channel quark exchange
and u-channel squark exchange interactions. In order to take
into consideration the polarization  in the calculation of the
differential cross-section  we use the density matrices of the
colliding beams. For polarized and unpolarized cases the density matrices of
photon is given as follows:

\begin{eqnarray}
\rho _{\mu \mu ^{\prime }}^{(\gamma )} &=&-\frac 12g_{\mu \mu ^{\prime }}%
\,\,\,\,\,\,\,\,\,\,\,\,\,\,\,\,\,\,\,\,\,\,\,\,\,\,\,\,\,\,\,\,\,\,\,\,\,\,\,\,%
\,\,unpolarized\,\,case
\end{eqnarray}

\begin{eqnarray}
\rho _{\mu \mu ^{\prime }}^{(\gamma )} &=&\frac 12(1+\overrightarrow{\xi }%
\cdot \overrightarrow{\sigma })_{ab}e_\mu ^{(a)}e_{\mu ^{\prime }}^{(b)}%
\,\,\,\,\,\,\,\,polarized\,\,case
\end{eqnarray}
where $\overrightarrow{\sigma }=(\sigma _1,\sigma _2,\sigma _3)$ are the usual
Pauli matrices and $\overrightarrow{\xi }=(\xi _1,\xi _2,\xi _3)$
are Stokes parameters of the backscattered laser photon. In our calculations
we take into account only circular polarization which is determined by
$\xi_2$. For the right (left) circular polarization, $\xi _2$ takes the value
of $+1(-1)$. $e_{\mu}^{(a)}$ $(a=1,2)$ are the polarization unit 4-vectors
which are orthogonal to each other and to the momenta of the colliding
particles. Density matrix for the quarks can be taken as in the form for the
massless spin 1/2 particles since the mass of quark has been ignored in
the calculation of the invariant amplitude.

\begin{eqnarray}
\rho ^q=\frac 12\gamma \cdot p(1\pm 2\lambda _q\gamma ^5)
\end{eqnarray}
Here $\lambda_{q}$ refers to the helicities of the quarks inside
the proton and takes the values of $+1/2(-1/2)$ for the positive
(or negative) helicity corresponding to the spin direction to the paralel
(or antiparallel) of its momentum. The calculation of differential
cross-section has been performed in the center of mass frame. One can easily
obtain the total cross-section $\hat{\sigma}$ for the subprocess
under consideration by integrating over $\hat{t}$:
\begin{eqnarray}
\hat{\sigma }(m_{\tilde{g}},m_{\tilde{q}%
},\hat{s},\xi _2,\lambda _q)&=&C\frac 12(1+2\lambda _q%
)\left[ A(m_{\tilde{g}},m_{\tilde{q}},\hat{s})+\xi _2B(m_{%
\tilde{g}},m_{\tilde{q}},\hat{s})\right]
\end{eqnarray}
here $C$ is a coefficient which includes coupling constants and
color factor, $\xi_2$ is the helicity for the backscattered laser photon.
For the subprocess cross-section $\hat{\sigma}(m_{\tilde{q}},m_{\tilde{g}%
},\hat{s},\xi_{2},\lambda_{q})$ we can use the following short notations:

\begin{eqnarray}
\hat{\sigma }(\xi _2 &=&+1,\lambda _q=+\frac 12)\equiv
\hat{\sigma }_{++}\nonumber\\
\hat{\sigma }(\xi _2 &=&-1,\lambda _q=+\frac 12)\equiv
\hat{\sigma }_{-+}\nonumber\\
\hat{\sigma }(\xi _2 &=&+1,\lambda _q=-\frac 12)\equiv
\hat{\sigma }_{+-}=0\nonumber\\
\hat{\sigma }(\xi _2 &=&-1,\lambda _q=-\frac 12)\equiv
\hat{\sigma }_{--}=0
\end{eqnarray}
From Eq.(19) it can be easily seen that the last two relations in Eq.(20)
vanish for $\lambda_{q}=-1/2$. In order to obtain the total cross-section
for the process $\gamma p\rightarrow \tilde{q}~\tilde{g}~X$ one should
perform the integration over the quark and photon distributions.
After making the change of the variables as in the previous section
(here $\tau _{\min }=(m_{\tilde{g}}+m_{\tilde{q}})^2/s$) one
 can write the total cross-section for the right circular
polarized laser and spin-parallel proton beam polarized longitudinally
as follows:

\begin{eqnarray}
\sigma _{R\uparrow } &=&\int_{\tau _{\min }}^{0.83}d\tau \int_{\tau /0.83}^1%
\frac{dx}x\left\{ P\left[ f_{+}^\gamma (\frac \tau x)f_{+/\uparrow }^q(x)%
\hat{\sigma }_{++}+f_{+}^\gamma (\frac \tau x)f_{-/\uparrow
}^q(x)\hat{\sigma }_{+-}\right] \right.  \nonumber \\
&&+\left. (1-P)\left[ f_{+}^\gamma (\frac \tau x)f_{+/\downarrow }^q(x)%
\hat{\sigma }_{+-}+f_{+}^\gamma (\frac \tau x)f_{-/\downarrow
}^q(x)\hat{\sigma }_{++}\right] \right\}
\end{eqnarray}
where P is the polarization percentage of spin-parallel protons in the beam
and  $f_{\pm}^{\gamma}(y)$ is the energy spectrum of the laser photons
which is given as in Eqs.(1-2). In the numerical integration the helicity
of the backscattered laser
photon, $\lambda_{\gamma}$ is taken as in Eq.(3).

The total cross-section expression in Eq.(21) can be rearranged by
considering the distribution of the valance quark (u-type) inside the
proton and written in the form of:

\begin{eqnarray}
\sigma _{R\uparrow }=\int_{\tau _{\min }}^{0.83}d\tau \int_{\tau /0.83}^1%
\frac{dx}x\left\{ f_{+}^\gamma (\frac \tau x)u_v^{+}(x)\hat{%
\sigma }_{++}\right\}
\end{eqnarray}
where $u_{v}^{+}$, u-type valance quark distribution with
the positive helicity, is defined by the sum of the
nonpolarized and difference polarized quark distributions

\begin{eqnarray}
u_v^{+}(x)=\frac 12(u_{np}+\bigtriangleup u_{pol}) .
\end{eqnarray}
After inserting this expression into the Eq.(22) then we arrive at:

\begin{eqnarray}
\sigma _{R\uparrow }=\int_{\tau _{\min }}^{0.83}d\tau \int_{\tau /0.83}^1%
\frac{dx}x\left\{ f_{+}^\gamma (\frac \tau x)\frac 12(u_{np}+\bigtriangleup
u_{pol})\hat{\sigma }_{++}\right\} .
\end{eqnarray}
For the polarized and unpolarized valance quark distributions in the above
expression, the following relations have been used \cite{eich84}.

\begin{eqnarray}
u_{np}(x)&=&2.751x^{-0.412}(1-x)^{2.69}\\
\bigtriangleup u_{pol}(x)&=&2.139x^{-0.2}(1-x)^{2.4} .
\end{eqnarray}
After the numerical calculations we get the production cross sections
for the gluino-squark process and show the dependence of the total cross
sections on the masses of the SUSY particles for various proposed
$\gamma p$ colliders in Figs.4(a-c) and Figs.5(a-c).
The upper mass limits for SUSY particle can easily be found from the
figures by using the luminosities given in Table 2. These
values are calculated by taking $100$ events per running year as the
observation limit for SUSY particle and tabulated in the same table.

Furthermore it is possible to look at the polarization asymmetries
as a function of the sparticle mass which can be useful for determination
of the mass parameter. Asymmetry with respect to the polarization cases of
the laser beam is defined as in Eq.(14).
One can also consider another asymmetry defined using the results
of the up and down polarized proton beams defined as before
in Eq.(15).
In Figs.6(a-c) we have presented the results of the polarization
asymmetry with respect to the left-right polarized laser beams.\\
{\bf {\it Signature for gluino-squark production:}} \rm In order to
maintain baryon and lepton number conservation one usually
imposes R-parity in the SUSY models. If R-parity exactly conserved SUSY
particles can only be produced in pairs and also processes give large
missing energy. Usually the photino and sneutrino are taken as the lightest
SUSY particles and will not be observed. The possible decay modes
of squarks and gluinos depend on the mass spectrum and on the
coupling constants. The gluino will decay into a squark and antiquark.
The squark will mainly decay into a quark and a photino. There are
also possible decays of the squark into a wino or a zino with the
less branching ratios. One possibility of the decay of zino is the
decay into a neutrino and a sneutrino.
According to the results of decay channels, the signature
for the process $\gamma p\rightarrow\tilde{q}\tilde{g}X$ might be
in general multijets + large missing energy and missing $p_T$.
The definite polarization asymmetries associated
with the missing energy and momentum may help in separating these
events from the backgrounds.

In conclusion, our analysis shows that the future $\gamma p$ colliders
can have considerable capacities in addition to the well known $pp$ and
$e^+e^-$ colliders in the investigation of supersymmetric particles.

\section{Squark-chargino Production}

The subprocess contributing to our physical process ${\gamma}p\to
\tilde{w}\tilde{q}$X  is  ${\gamma}q\to\tilde{w}\tilde{q}$.
The invariant amplitude for the specific subprocess ${\gamma}u\to
\tilde{w}^{+}\tilde{d}$ is the sum of the three terms corresponding
to the s-channel $u$ quark exchange, the t-channel $\tilde{w}$ wino
exchange and the u-channel $\tilde{d}$ squark exchange interactions:
\begin{eqnarray}
{\cal M}_a &=&\frac{-iee_q g}{2\hat{s}}\bar{u}(p^{\prime})(1-\gamma_5)
(\not\! p+\not\! k)\not\! \epsilon u(p)\nonumber\\
{\cal M}_b &=&\frac{-iege_{\tilde{w}}}{2(\hat{t}-m^2_{\tilde{w}})}
\bar{u}(p^{\prime})\not\! \epsilon (\not\! p-\not\! k+m_{\tilde{w}})
(1-\gamma_5)u(p)\\
{\cal M}_c &=&\frac{-iee_{\tilde{q}}g}{2(\hat{u}-m^2_{\tilde{q}})}
\bar{u}(p^{\prime})(1-\gamma_5)u(p)(p-p^\prime +k^\prime ).\epsilon\nonumber
\end{eqnarray}
where $e_q$, $e_{\tilde{q}}$ and $e_{\tilde{w}}$ are the quark,
squark and wino charges, and $g=e/sin{\theta_{w}}$ is the weak coupling
constant. Note that we ignore the quark masses.

Since we are interested in the polarized cross-section, we use the following
density matrices for the initial photon and quark:

\begin{eqnarray}
\rho^{(\gamma)}&=&\frac{1}{2}(1+\vec{\xi}.\vec{\sigma})\nonumber\\
\rho^{(q)}&= &u\bar{u} =\not\! p[1+\gamma_{5}(\lambda_{q}+\vec{\xi}_{\perp}.
\vec{\gamma}_{\perp})]
\end{eqnarray}
where ${\xi}_1$, ${\xi}_2$ and ${\xi}_3$ are Stokes parameters. We take into
account only circular polarization for the photon which is defined
by ${\xi}_2$, as has been already mentioned in the previous section.
$\lambda_q$ stands for the helicity of the parton-quark that is
$+1 (-1)$ for the spin directions parallel (anti-parallel) to its
momentum. The last term in the quark density matrix does not contribute,
because after the integration over the azimuthal angle it vanishes.

One can easily obtain the differential cross section for the subprocess
${\gamma}u\to\tilde{w}^+\tilde{d}$  as follows
\begin{eqnarray}
\frac{d\hat{\sigma}}{d\hat{t}}=\frac{\pi \alpha^2}{\hat{s}^2 sin^2
\theta_w}(1+\lambda_{q})\biggr[\frac{d\hat{\sigma_0}}{d\hat{t}}
+\lambda(y)\frac{d\hat{\sigma}_1}{d\hat{t}}\biggr].
\end{eqnarray}
Performing the $d\hat{t}$ integration from $t_{min}$ to $t^{max}$ which
are given by
\begin{eqnarray}
t^{max}_{min}=\frac{1}{2}(m^2_{\tilde{w}}+m^2_{\tilde{q}}-\hat{s})
\biggr[1\mp\sqrt{1-4m^2_{\tilde{w}}m^2_{\tilde{q}}/(m^2_{\tilde{w}}
+m^2_{\tilde{q}}-\hat{s})^2}\biggr]
\end{eqnarray}
we immediately get the total cross section as
\begin{eqnarray}
\hat{\sigma} (m_{\tilde{w}},m_{\tilde{q}},\hat{s},\lambda{(y)})
=\frac{\pi \alpha^2}{\hat{s}^2 sin^2 \theta_w}(1+\lambda_q)
\biggr[\hat{\sigma_{0}}(m_{\tilde{w}},m_{\tilde{q}},\hat{s})
+{\lambda(y)} \hat{\sigma_{1}}(m_{\tilde{w}},m_{\tilde{q}},\hat{s}) \biggr].
\end{eqnarray}
Note that the cross sections ( Eqs(29) and (31)) are zero for $\lambda_{q}
=-1$ because of the fact that we ignore the quark mass. Integrating the
subprocess cross-section $\hat{\sigma}$ over the quark and photon distributions
we obtain the total cross-section for the physical process
$\gamma p\to\tilde{w} \tilde{q}X$:

\begin{eqnarray}
\sigma=\int^{0.83}_{(m_{\tilde{w}}+m_{\tilde{q}})^{2}/s} d\tau
\int^{1}_{\tau/0.83} \frac{dx}{x} f_{\gamma /e}(\tau/x) f_{q}(x)
\hat{\sigma}(m_{\tilde{w}},m_{\tilde{q}},\hat{s},\lambda(\tau/x))
\end{eqnarray}
where the photon distribution function, $f_{{\gamma}/e}(y)$, is actually the
normalized differential cross-section of the Compton backscattering, Eq.(1)
; $f_q (x)$ is the distribution of quarks inside the proton. We set
$\lambda_q=+1$ and $f_q (x)\to u^{+} (x)=\frac{1}{2}(u_{unp}+{\Delta}u_{pol})$
for the $u$-type valence quark distribution. In our numerical calculations,
we use the distribution functions given in Eqs.(25-26).
Performing the integrations in Eq.(32) numerically we obtain
the total cross-section for the associated wino-squark production.
We plot the dependence of the total cross-sections on the masses of
the SUSY particles for various proposed $\gamma p$ colliders in
Figs. 7(a-c) for $\lambda_0 \lambda_e =+1$ and $-1$.
By taking 100 events per running year
as observation limit for a SUSY particle, one can easily find the upper
discovery mass limits from these figures using the luminosities of
the proposed $\gamma p$ colliders given in Table 3. These discovery
limits are tabulated in the same table.

It may be more interesting to use a polarization asymmetry in determining
the masses of SUSY particles. Such an asymmetry can be defined with
respect to the product of the polarizations of the laser photon and
the electron as follows
\begin{eqnarray}
A=\frac{\sigma_{-}-\sigma_{+}}{\sigma_{-}+\sigma_{+}}
\end{eqnarray}
where $\sigma_{+}$ and $\sigma_{-}$ are the polarized total cross-sections
given in Figs. 7(a-c).
The results of the polarization asymmetry are shown in Figs. 8(a-c) for
three colliders.
{\bf {\it Signature for the squark-chargino production:}} \rm If one
compares the curves $\sigma_{+}$ and $\sigma_{-}$ in Figs. 7(a-c)
for each collider one sees that the
polarized cross-sections for different polarization do not differ much
from each other and also from the unpolarized ones. Therefore, the discovery
mass limits for SUSY partners obtained with polarized beams are nearly
equal to those obtained with unpolarized beams. But the polarization
asymmetry is highly sensitive to the wino
and the squark masses and as high as 0.4 for all cases.
Especially in the case of $m_{\tilde{q}}=250$ $GeV$,
the asymmetry parameter A is around 0.6 for the higher wino masses.

The signature of the associated $\tilde{w}^{+} \tilde{d}$ production
will depend on the mass spectrum of SUSY particles. It is generally assumed
that the photino and sneutrino are the lightest SUSY particles and
that the hierarchy of the squark masses is similar to that of quarks.
With these assumption we have the following decays for the case
$m_{\tilde{q}}=m_{\tilde{w}}$:

$\tilde{d}\to d\tilde{\gamma}$, $d\tilde{g}$   and
  $\tilde{w}\to l^{+}\tilde{\nu}$,  $\nu\tilde{l}^{+}$,  $W^{+}\tilde{\gamma}$

By taking into account the further decays $\tilde{l}^{+}\to l\tilde{\gamma}$
and $W^{+}\to l^{+}\nu, q \bar{q}$ we arrive at the ultimate final states
as

$l^{+}+n$ $jets(n=1,3,5)+$ large missing energy and missing $P_{T}$

The main background for the final state $l^{+}+jet+P_{T}^{miss}$ will come
from the process $\gamma q\to W q \to l^{+}\nu q$; but this background
may be reduced, in principle,by the cut $P_{T}^{miss}>45$ $GeV$ if
$m_{\tilde{w}} \gg m_{W}$.

\widetext \figure{(a-c) Production cross sections of squark pairs
as a function of their masses for various $\gamma p$ colliders. In
each figure the polarization states of the beams are indicated as
follows: Line: Unpolarized beams, linespoints: right polarized
laser, up protons, dots: right polarized laser, down
protons\label{fig1}}

\figure{(a-c) Up-down asymmetry versus squark masses for different
$\gamma p$ colliders.\label{fig2}}

\figure{Left-right asymmetry versus squark mass for the HERA+LC
$\gamma p$ collider.\label{fig3}}

\figure{(a-c) Production cross sections for gluino-squark as a
function of their masses for various colliders. Each figure
corresponds to right circular polarized laser and spin-parallel
proton beam polarized longitudinally.\label{fig4}}

\figure{(a-c) Production cross sections for gluino and squarks as
a function of their masses for various colliders. Each figure
corresponds to left circular polarized laser and spin-parallel
proton beam polarized longitudinally.\label{fig5}}

\figure{(a-c) Left-right polarization asymmetries versus sparticle
masses for different $\gamma p$ colliders.\label{fig6}}

\figure{(a-c) Squark-wino production cross-sections as a function
of the mass. Two thin curves on the left stand for the collider
HERA+LC, middle curves for LHC+Linac 1 and two curves on the right
for LHC+TESLA. A little bit higher (lower) curve of each twin is
for $\lambda_{0} \lambda_{e}=-1$ ($\lambda_{0} \lambda_{e}=+1$),
i.e., $\sigma_{-}$ ($\sigma_{+}$).\label{fig7}}

\figure{(a-c) Polarization asymmetries as a function of the mass.
Thin curve stands for the collider HERA+LC, middle curve for
LHC+Linac 1 and thick curve for LHC+TESLA.\label{fig8 }}

\begin{table}
\caption{Parameters of $\gamma p $ colliders  and discovery limits
for scalar quarks with $e_{\tilde{q}}=2/3$. (Note:$\sqrt{s_{\gamma
p}}^{max}=0.91 \sqrt{s_{ep}}$)\label{tab1}}
\begin{tabular}{lllll}
 Machines & $\sqrt{s_{ep}}$ & ${\cal L}_{\gamma p}$
& $m_{\tilde {q}}$ (TeV)  & $m_{\tilde{q}}$ (TeV) \\
           &(TeV) & ($10^{30} cm^{-2}s^{-1}$)
&  (non-degenerate)  & (degenerate) \\
           &   &    & squarks & squarks  \\
\tableline
HERA+LC    &  1.28   &  25      & 0.2      & 0.3  \\
 LHC+TESLA  &   5.55  &   500    &  0.85   & 1.1 \\
 LHC+e-Linac &    3.04 &   500    &  0.6   & 0.7 \\
\end{tabular}
\end{table}

\begin{table}
\caption{Parameters of different $\gamma p$ colliders  and
discovery mass limits for scalar quarks  and gluinos. (Note:
$\sqrt{s_{\gamma p}}\;^{max}=0.91 \sqrt{s_{ep}}$)\label{tab2}}
\begin{tabular}{|l|l|l|l|l|l|}
\hline
{\bf Machines} &
\begin{tabular}{l}
$\sqrt{s_{ep}}$ \\
$(TeV)$%
\end{tabular}
&
\begin{tabular}{l}
${\cal L}_{\gamma p}$ \\
$(10^{30} cm^{-2}s^{-1})$
\end{tabular}
&
\begin{tabular}{l}
$m_{\tilde{u}}=m_{\tilde{g}}$ \\
$(TeV)$%
\end{tabular}
&
\begin{tabular}{l}
$m_{\tilde{u}}=0.10$ \\
$m_{\tilde{g}}\quad (TeV)$
\end{tabular}
&
\begin{tabular}{l}
$m_{\tilde{g}}=0.10$ \\
$m_{\tilde{u}}\quad (TeV)$
\end{tabular}
\\ \hline
{\bf HERA+LC}     & $1.28$ & $25$  & $0.17$ & $0.32$ & $0.21$ \\ \hline
{\bf LHC+TESLA}   & $5.55$ & $500$ & $0.77$ & $1.90$ & $1.05$ \\ \hline
{\bf LHC+e-Linac} & $3.04$ & $500$ & $0.53$ & $1.22$ & $0.75$ \\ \hline
\end{tabular}
\end{table}

\begin{table}
\caption{Energy and luminosity values of different $\gamma$p
colliders. The discovery mass limits for squarks and winos are
given in last three columns for $\lambda_0 \lambda_e=+1$
($\lambda_0 \lambda_e=-1$).\label{tab3}}

\begin{tabular}{|l|l|l|l|l|l|}
\hline
{\bf Machines} &
\begin{tabular}{l}
$\sqrt{s_{ep}}$ \\
$(TeV)$%
\end{tabular}
&
\begin{tabular}{l}
${\cal L}_{\gamma p}$ \\
$(10^{30} cm^{-2}s^{-1})$%
\end{tabular}
&
\begin{tabular}{l}
$m_{\tilde{q}}=m_{\tilde{w}}$ \\
$(TeV)$%
\end{tabular}
&
\begin{tabular}{l}
$m_{\tilde{q}}=0.25$ \\
$m_{\tilde{w}}\quad (TeV)$%
\end{tabular}
&
\begin{tabular}{l}
$m_{\tilde{w}}=0.10$ \\
$m_{\tilde{q}}\quad (TeV)$%
\end{tabular}
\\ \hline
{\bf HERA+LC} & $1.28$ & $25$ &$0.23(0.25)$ & $0.18(0.23)$ & $0.35(0.40)$ \\ \hline
{\bf LHC+Linac 1} & $3.04$  & $500$ & $0.65(0.70)$ & $1.00(1.15)$ & $1.33(1.48)$ \\ \hline
{\bf LHC+TESLA} & $5.55$ & $500$ & $0.95(1.05)$ & $1.55(1.70)$ & $2.15(2.50)$ \\ \hline
\end{tabular}
\end{table}

\begin{center}
\epsfig{file=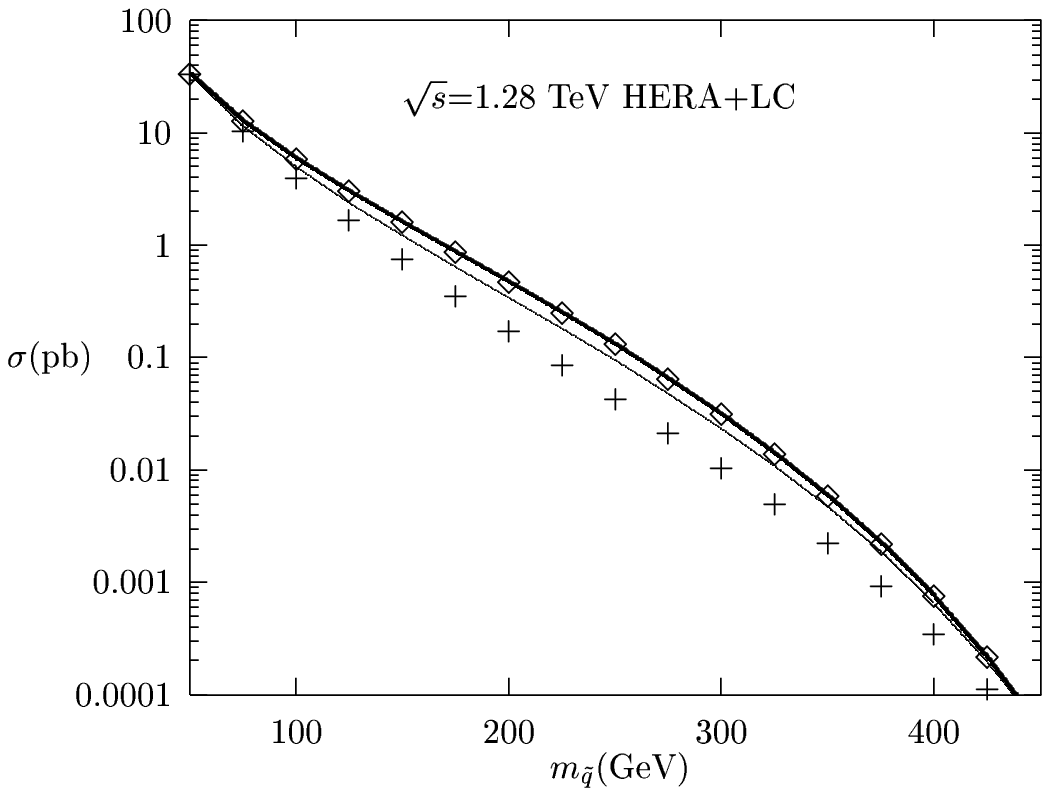}
\smallskip

Fig.1.a
\bigskip

\epsfig{file=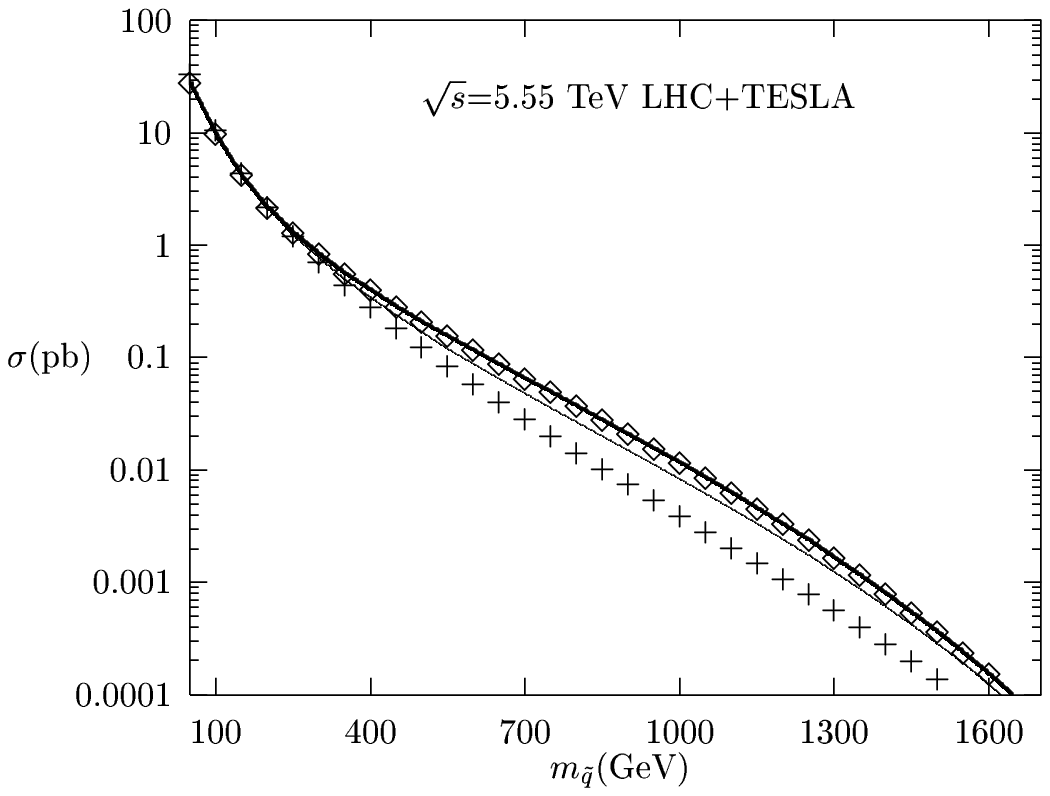}
\smallskip

Fig.1.b

\epsfig{file=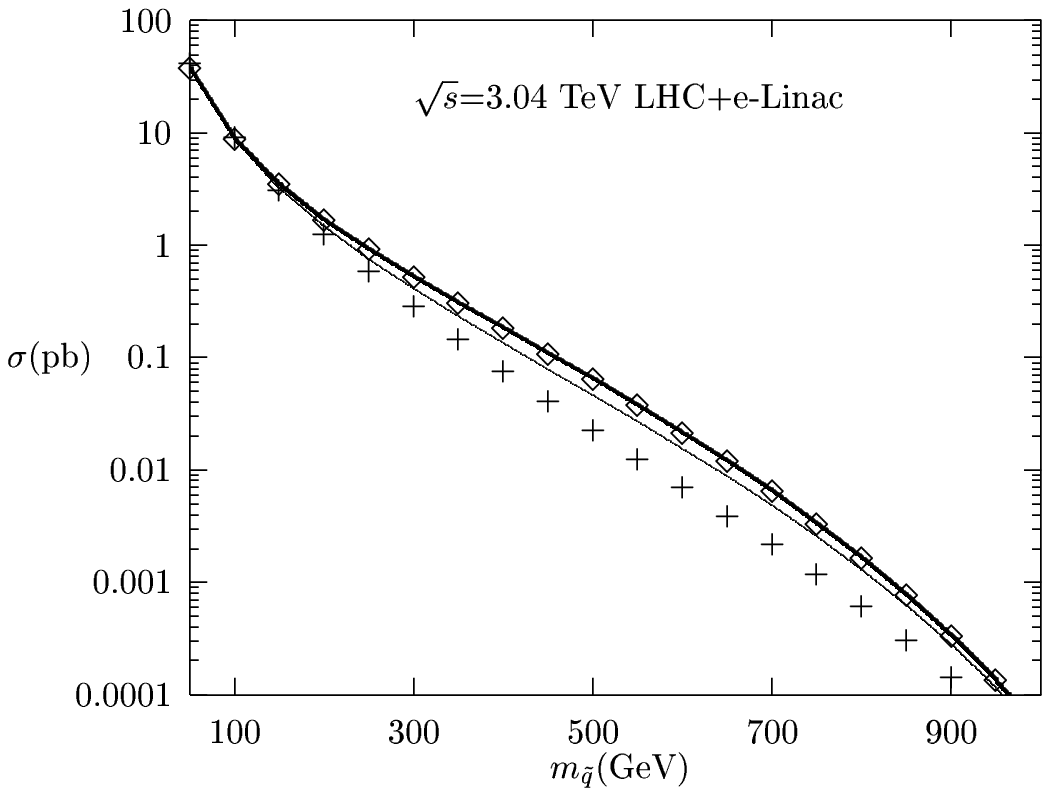}
\smallskip

Fig.1.c

\epsfig{file=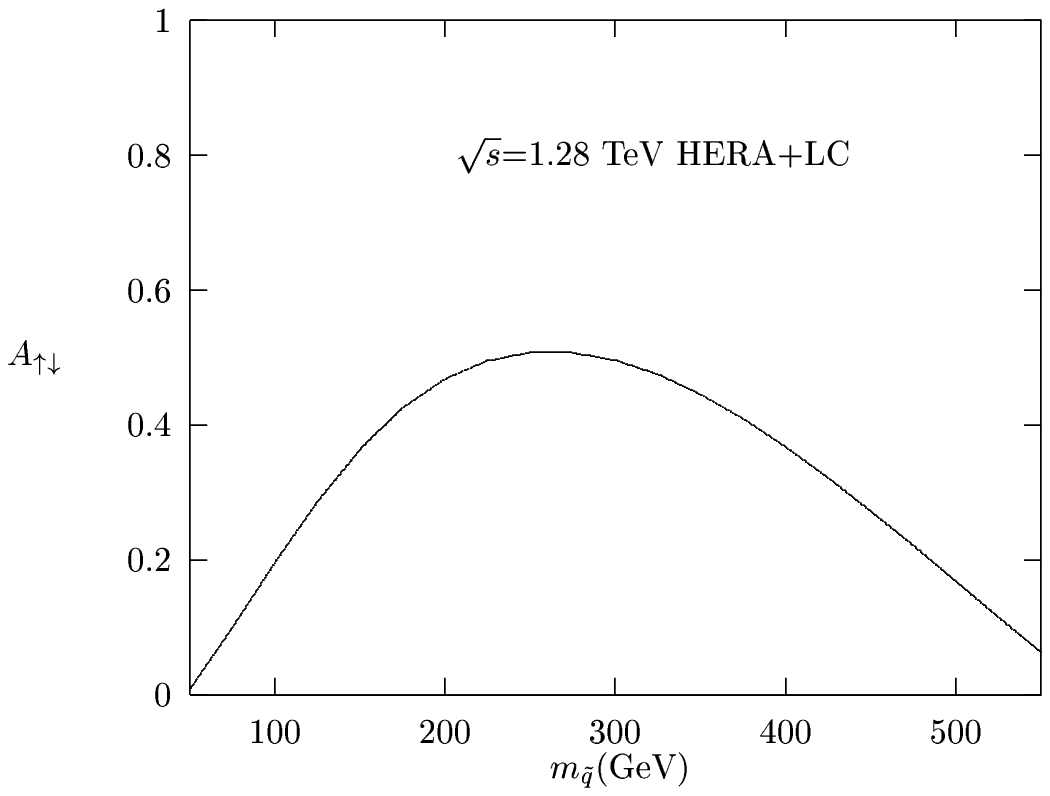}
\smallskip

Fig.2.a

\epsfig{file=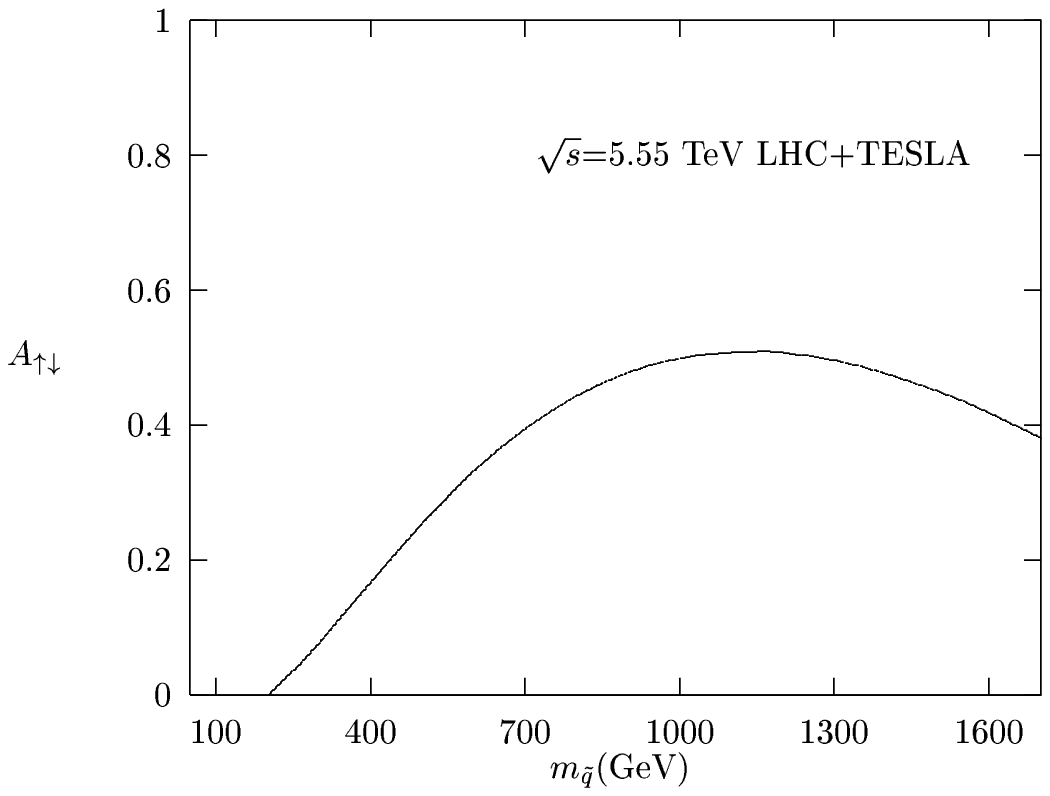}
\smallskip

Fig.2.b

\epsfig{file=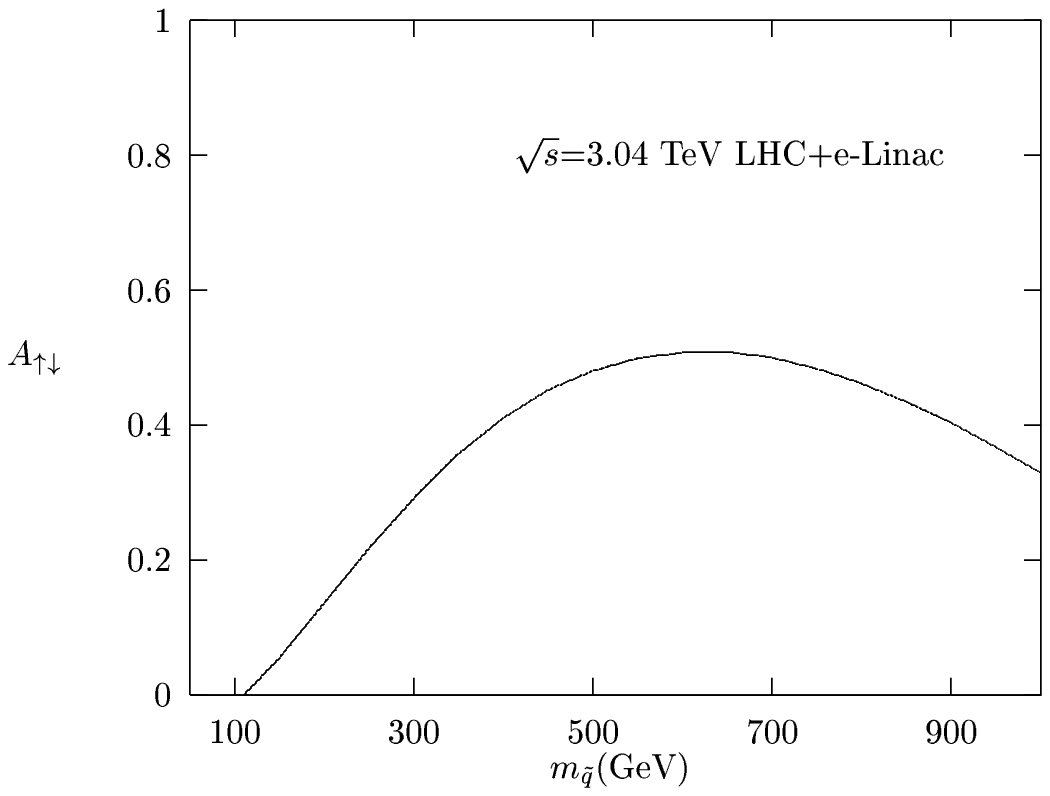}
\smallskip

Fig.2.c

\epsfig{file=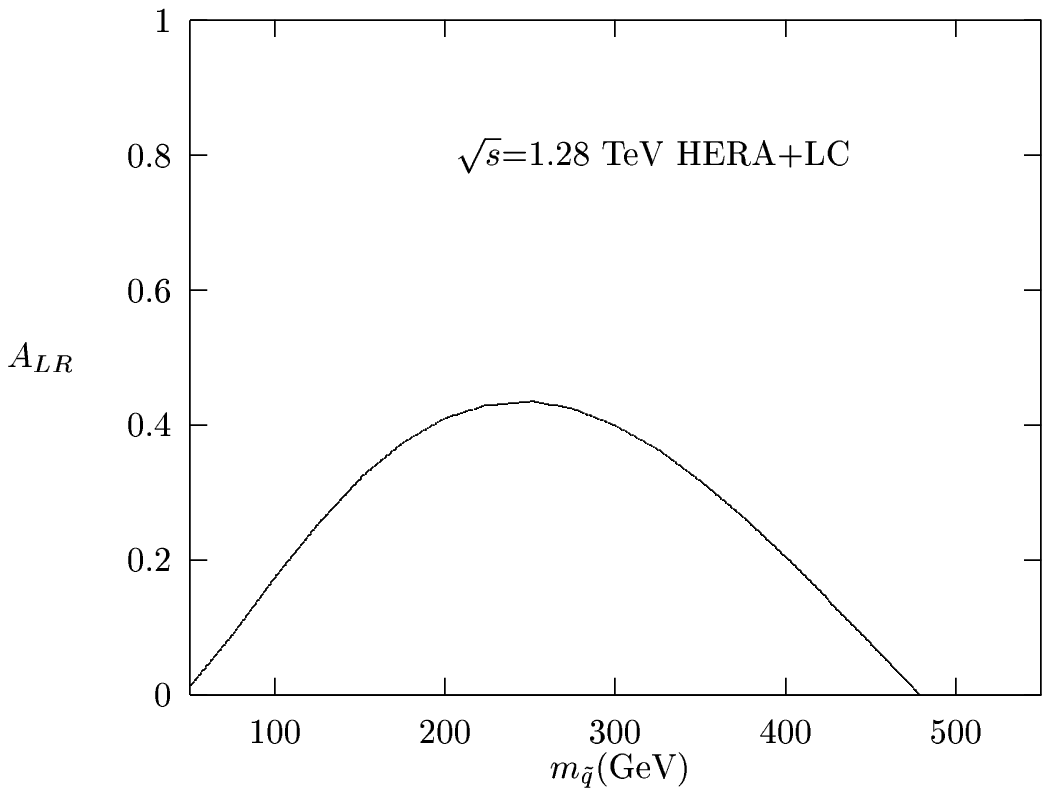}
\smallskip

Fig.3.b

\end{center}

\begin{center}
\epsfig{file=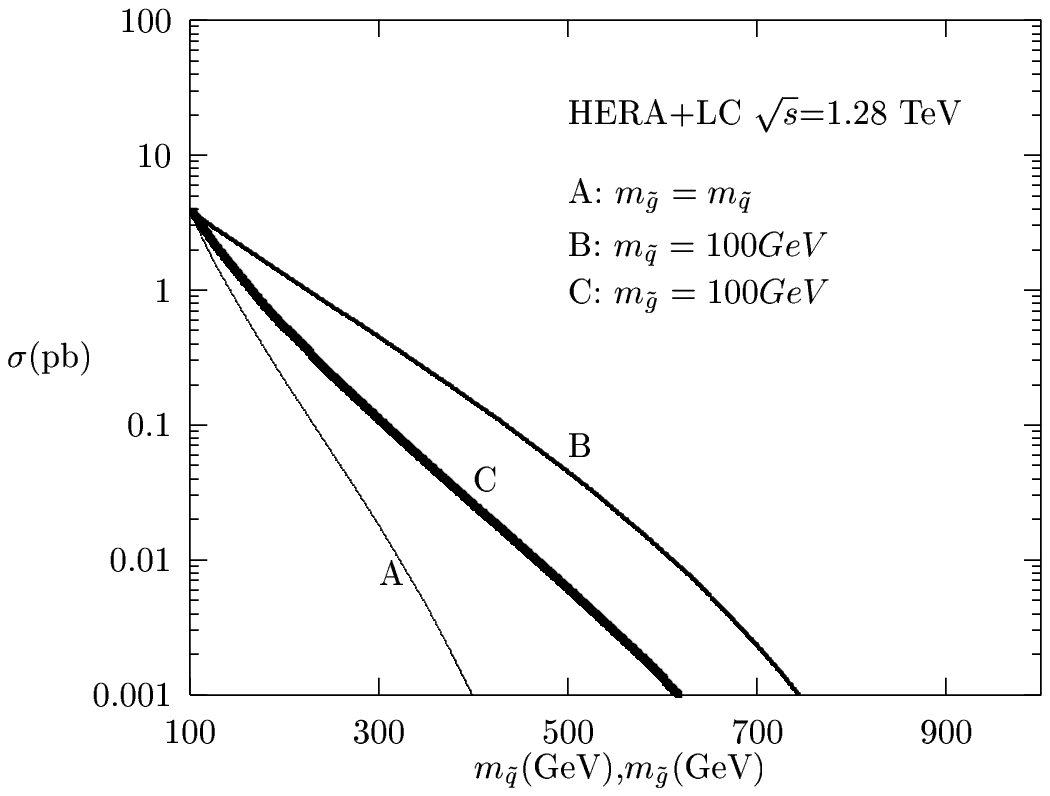}
\smallskip
\center{Fig.4.a}

\bigskip
\epsfig{file=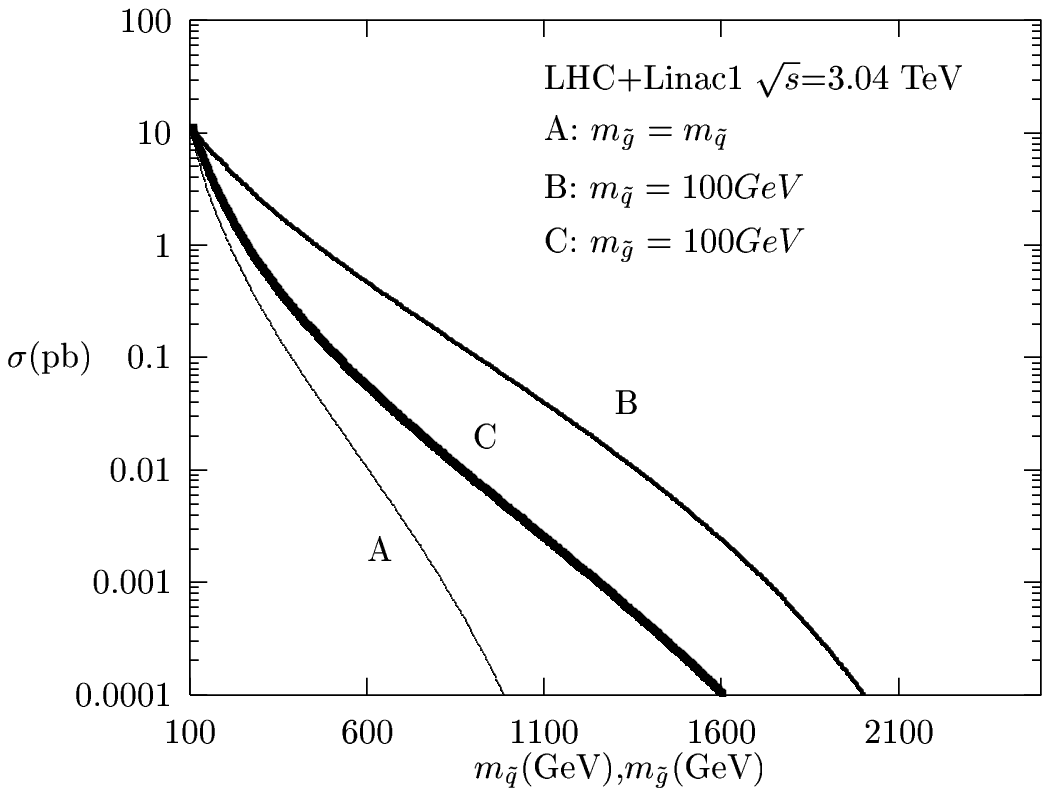}
\smallskip
\center{Fig.4.b}
\bigskip

\epsfig{file=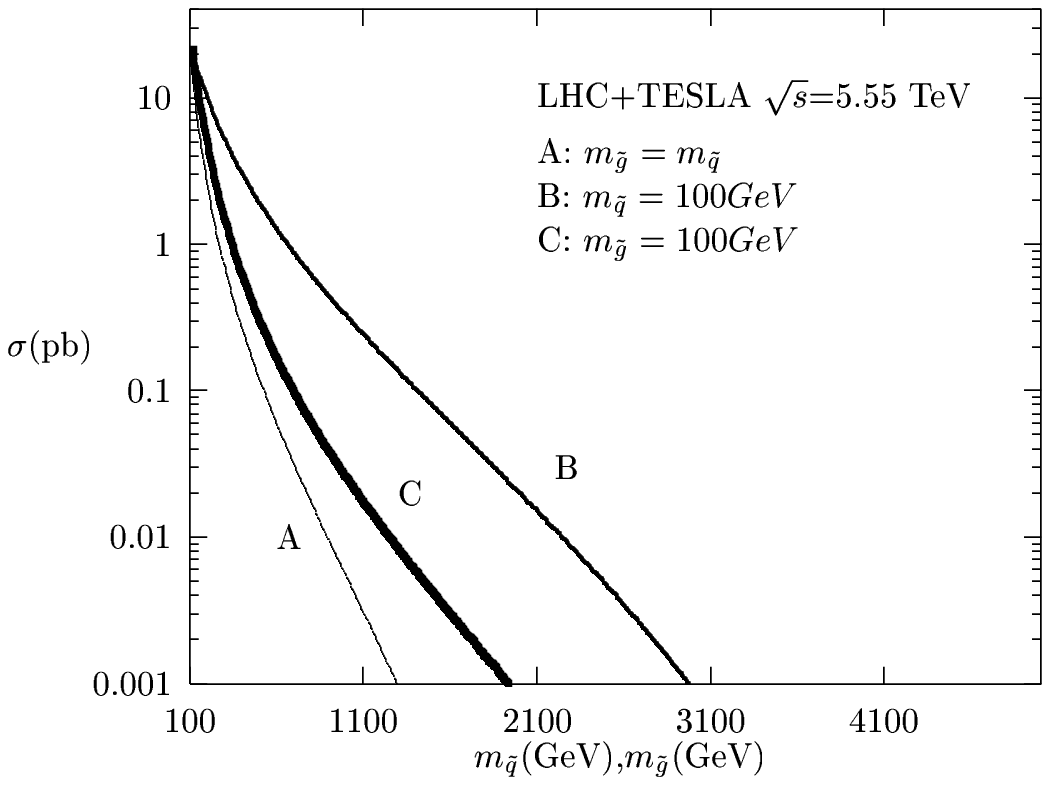}
\smallskip
\center{Fig.4.c}
\bigskip

\epsfig{file=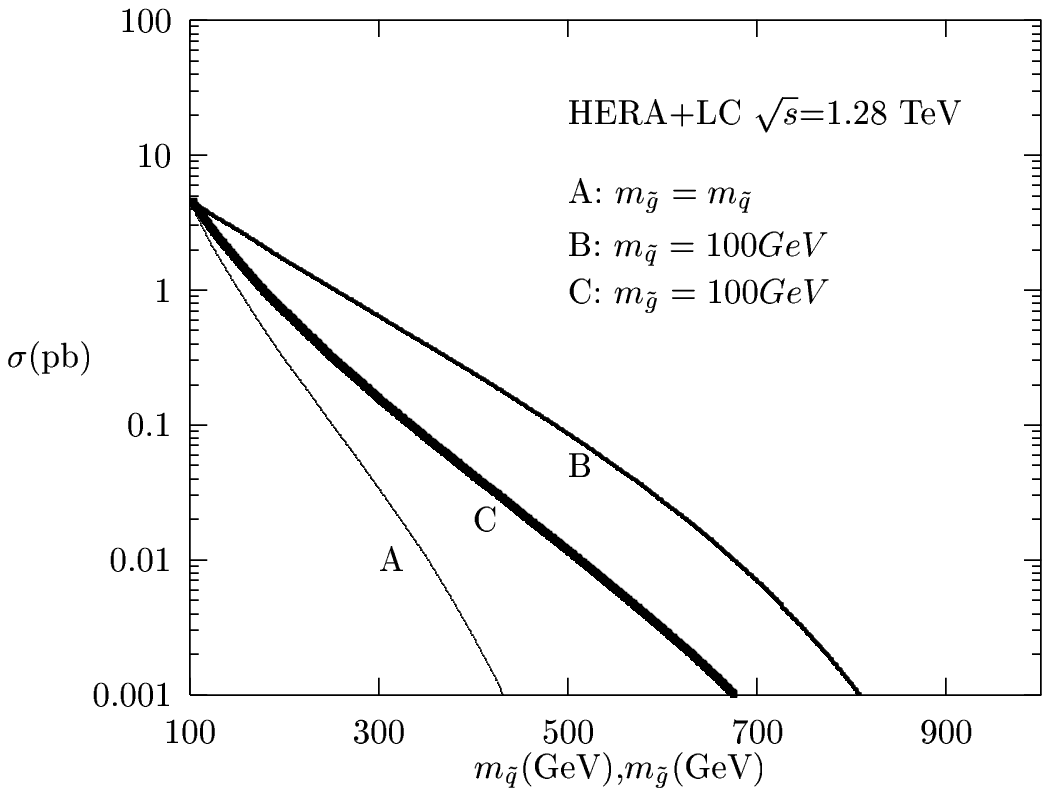}
\smallskip
\center{Fig.5.a}
\bigskip

\epsfig{file=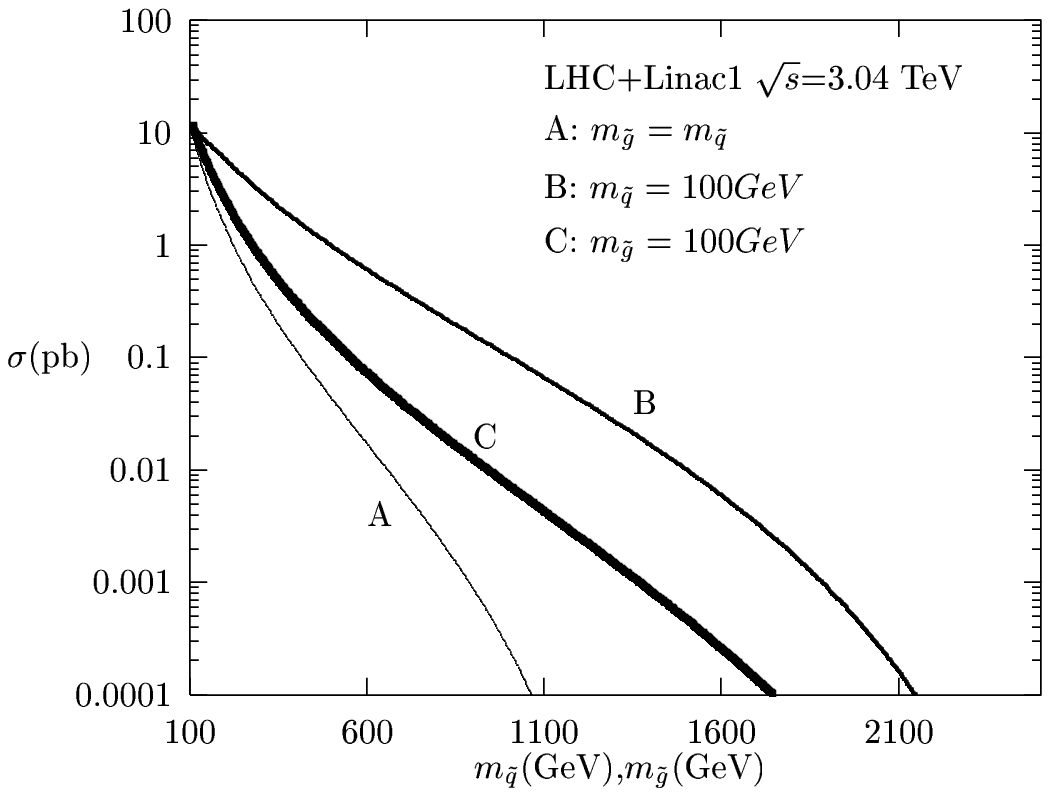}
\smallskip
\center{Fig.5.b}
\bigskip

\epsfig{file=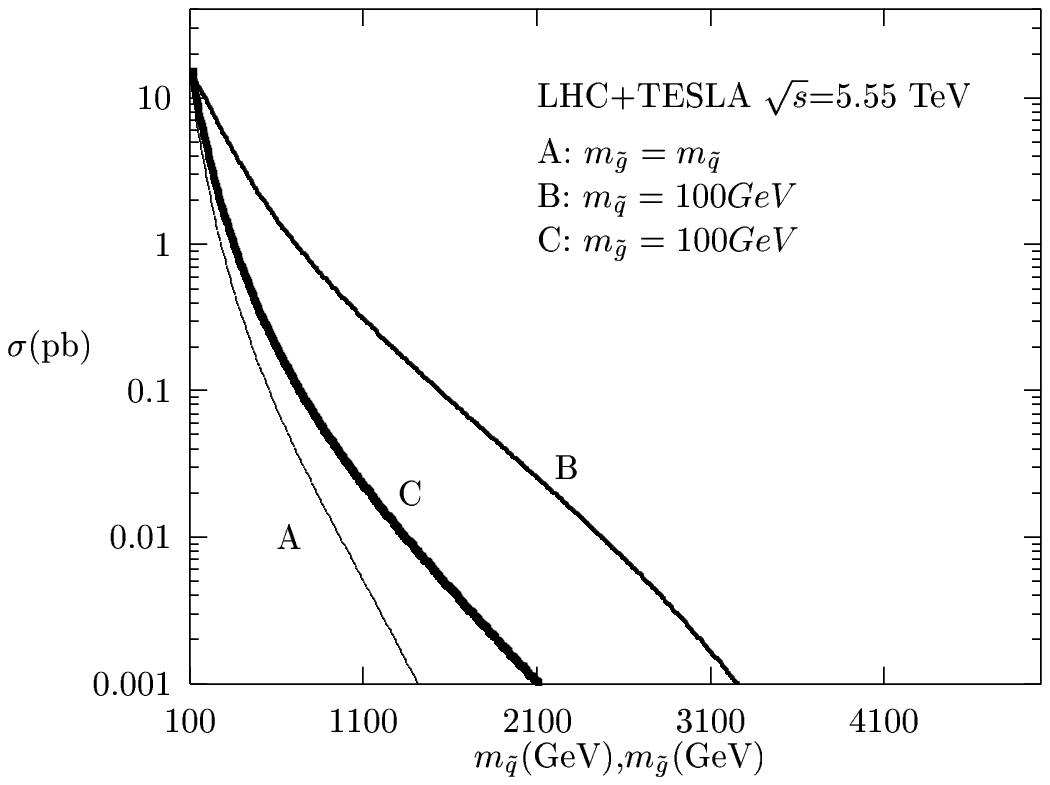}
\smallskip
\center{Fig.5.c}
\bigskip

\epsfig{file=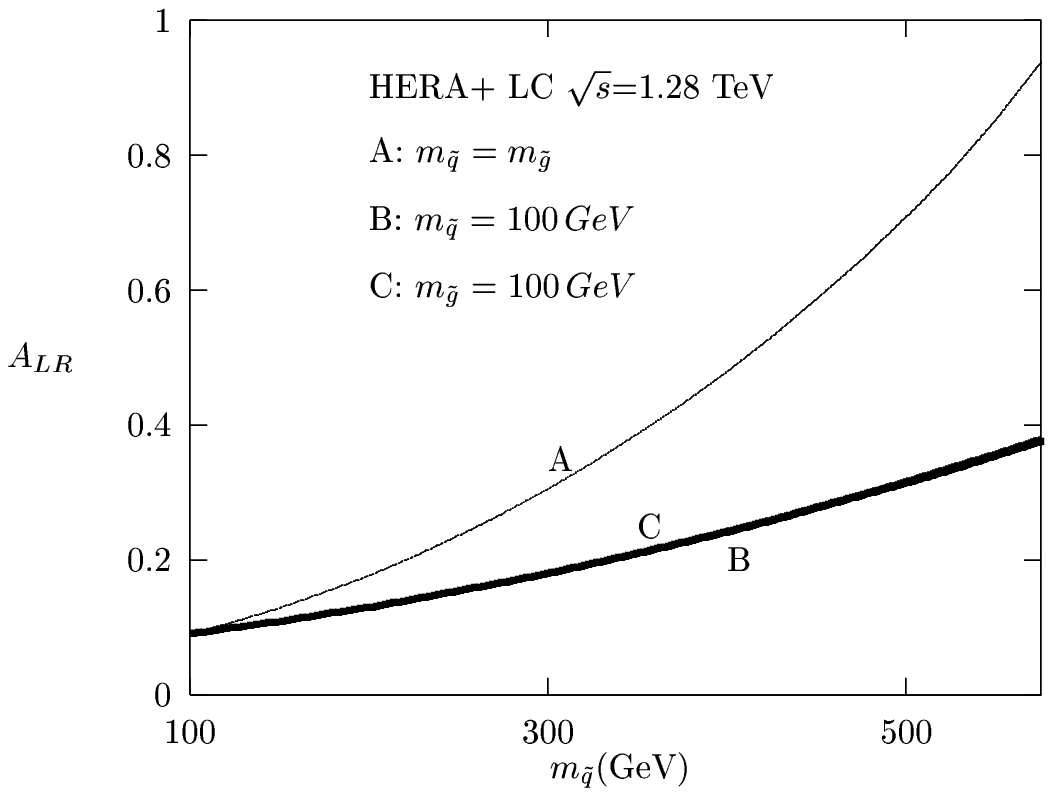}
\smallskip
\center{Fig.6.a}
\bigskip

\epsfig{file=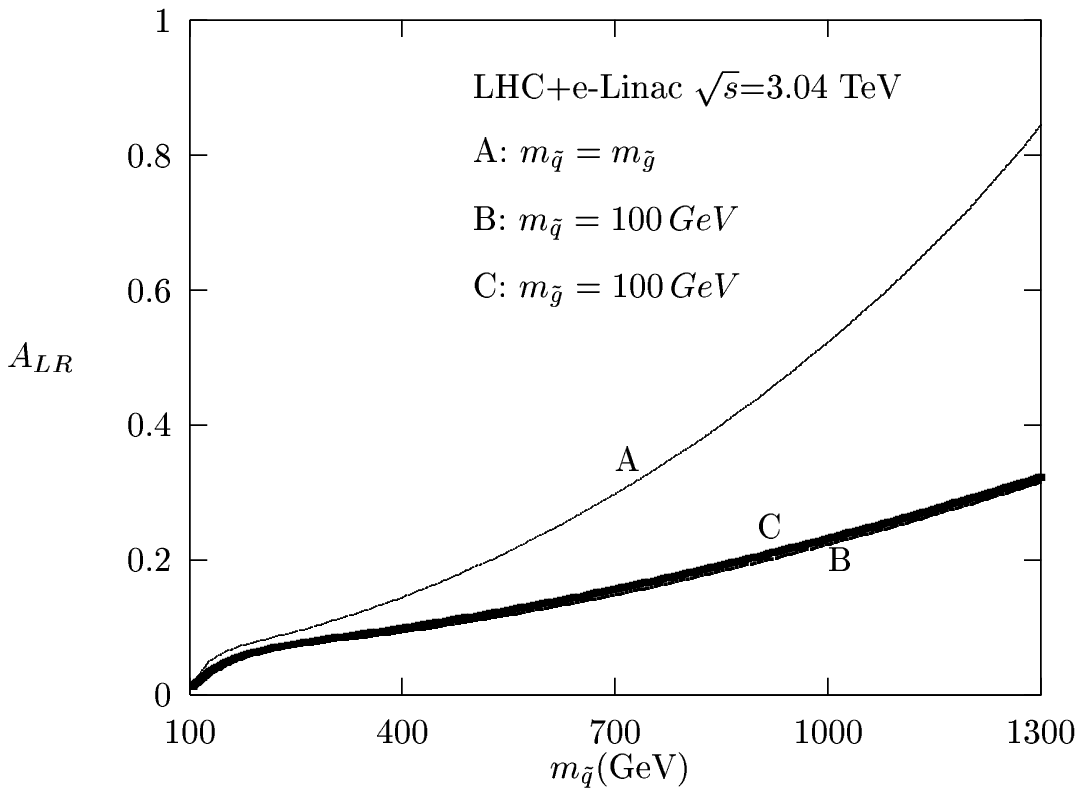}
\smallskip
\center{Fig.6.b}
\bigskip

\epsfig{file=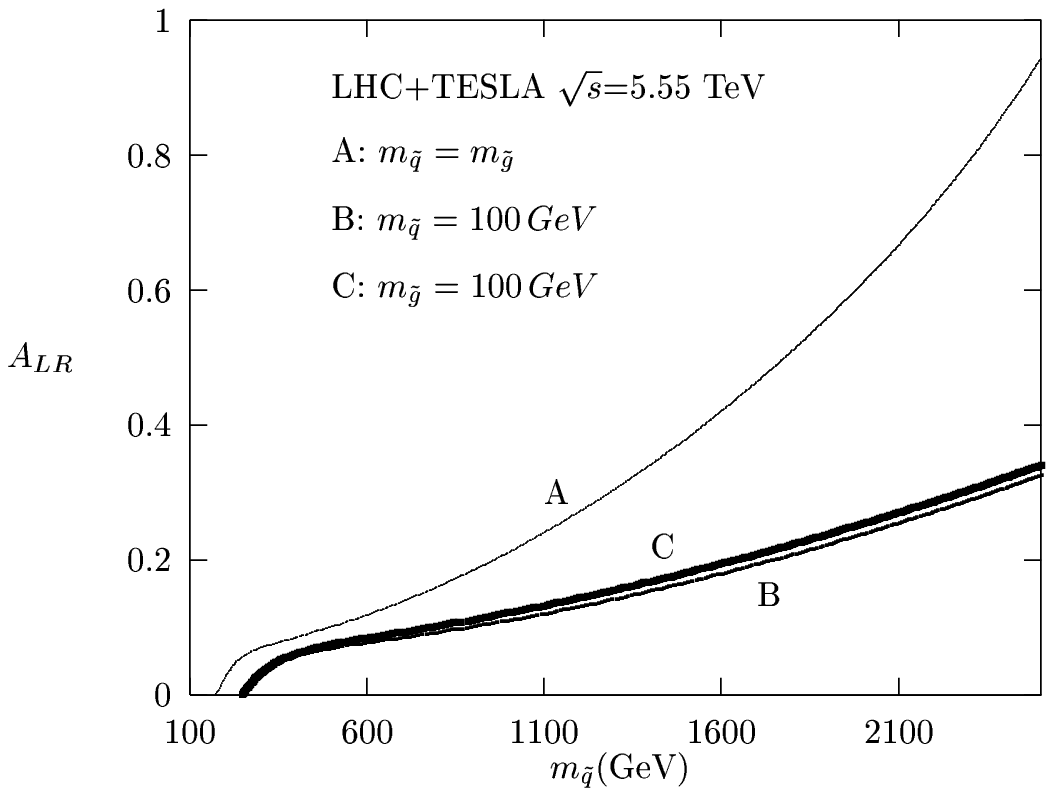}
\smallskip
\center{Fig.6.c}
\bigskip
\end{center}

\begin{center}
\epsfig{file=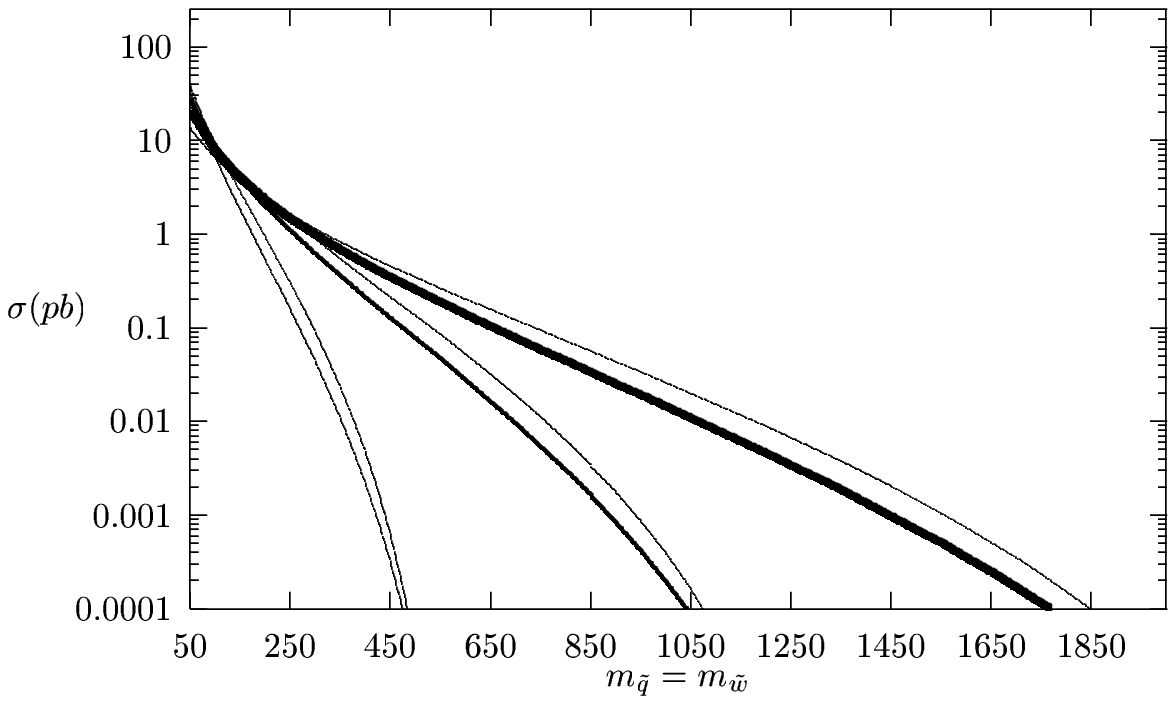}
\smallskip

Fig.7.a

\epsfig{file=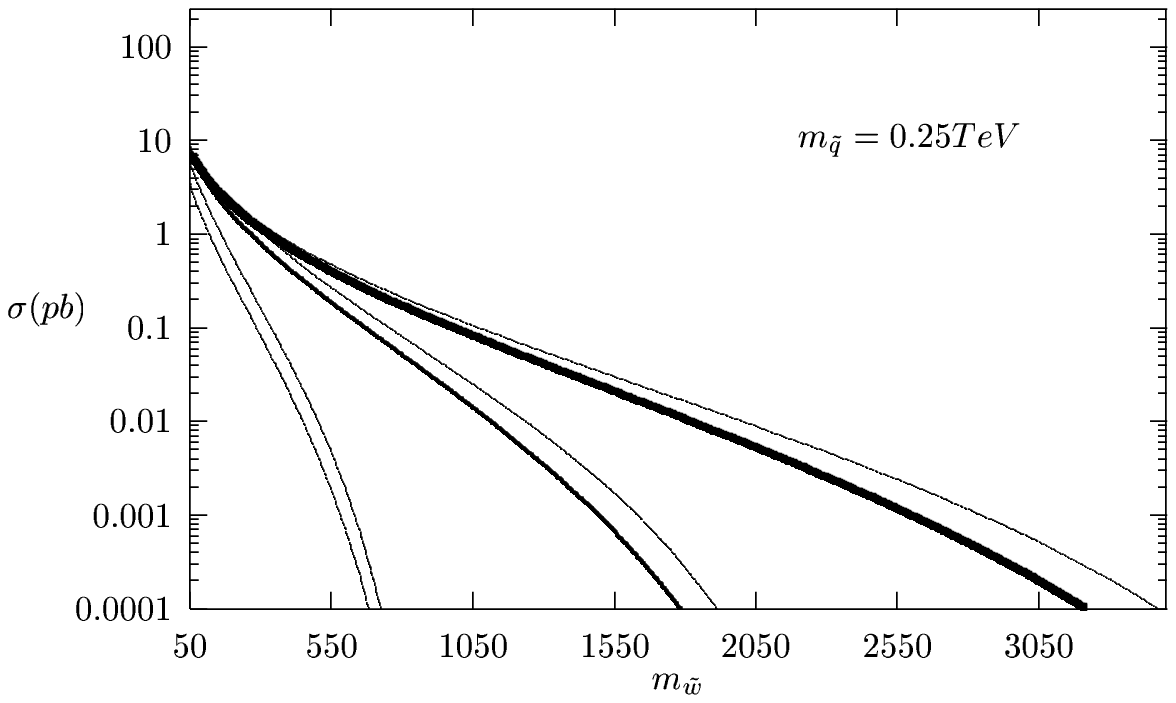}
\smallskip

Fig.7.b

\epsfig{file=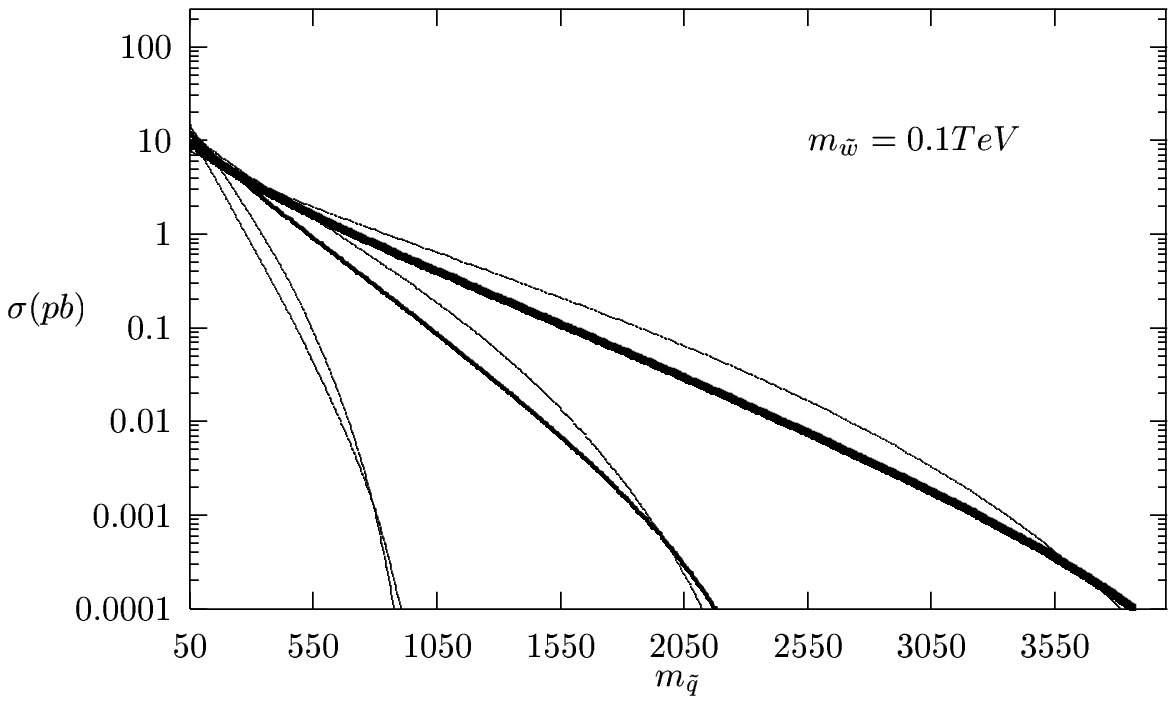}
\smallskip

Fig.7.c

\epsfig{file=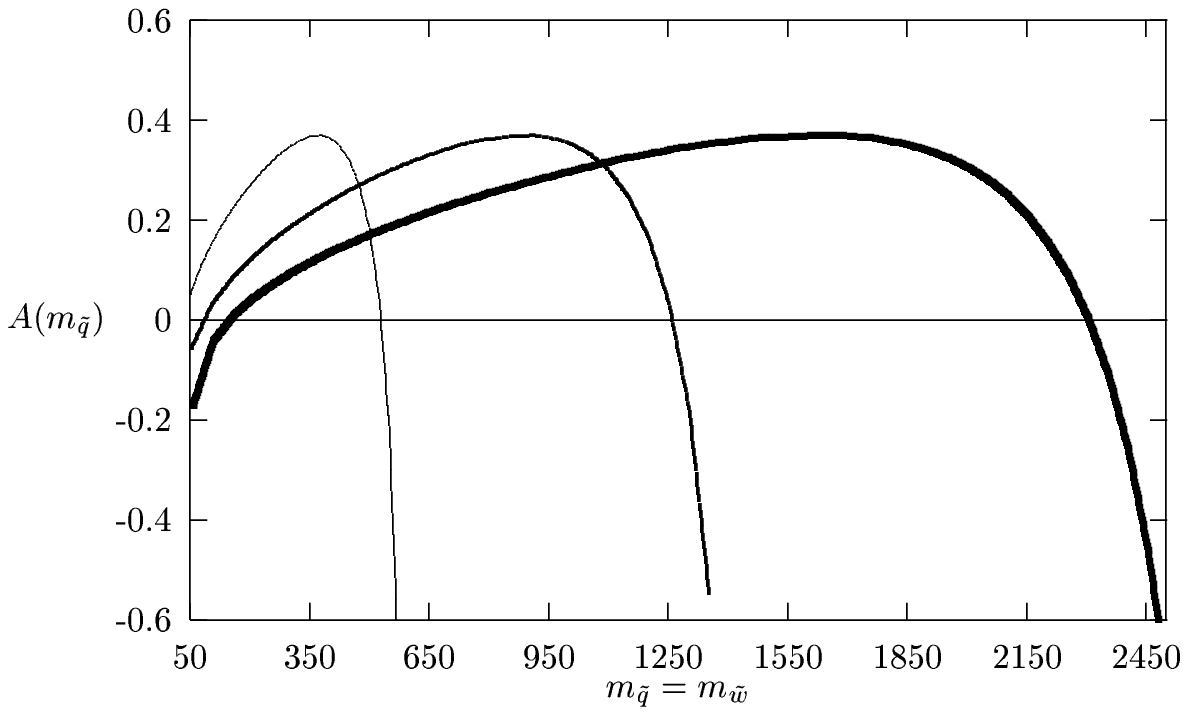}
\smallskip

Fig.8.a

\epsfig{file=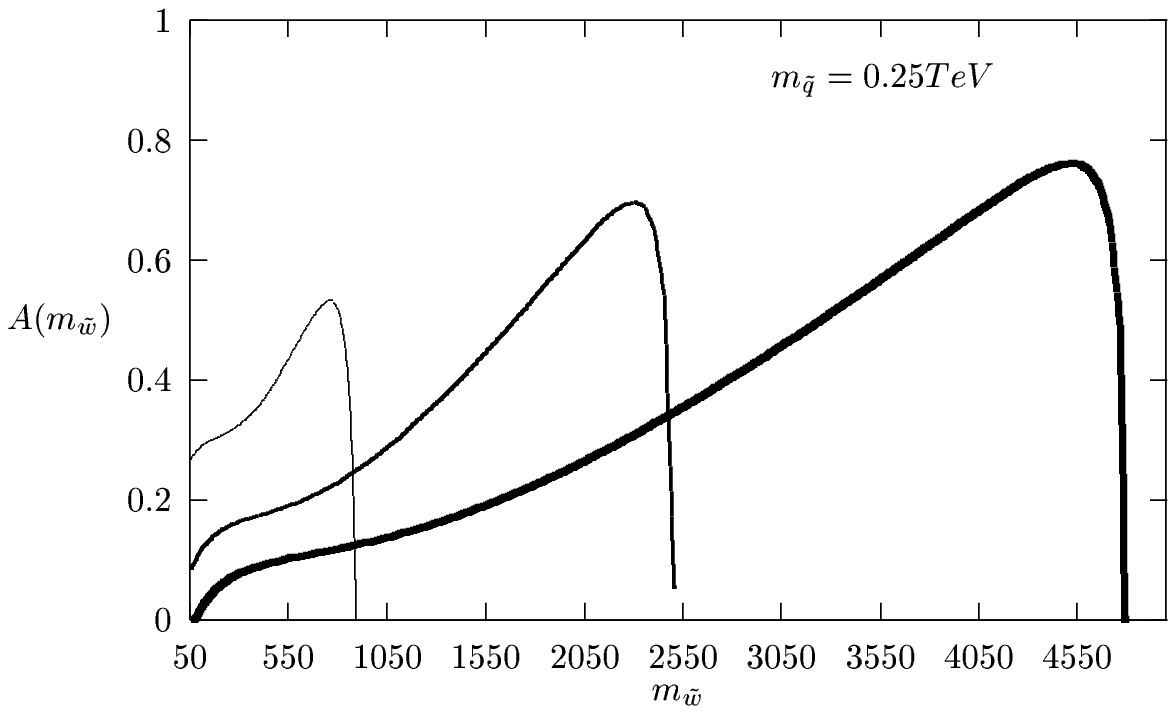}
\smallskip

Fig.8.b

\epsfig{file=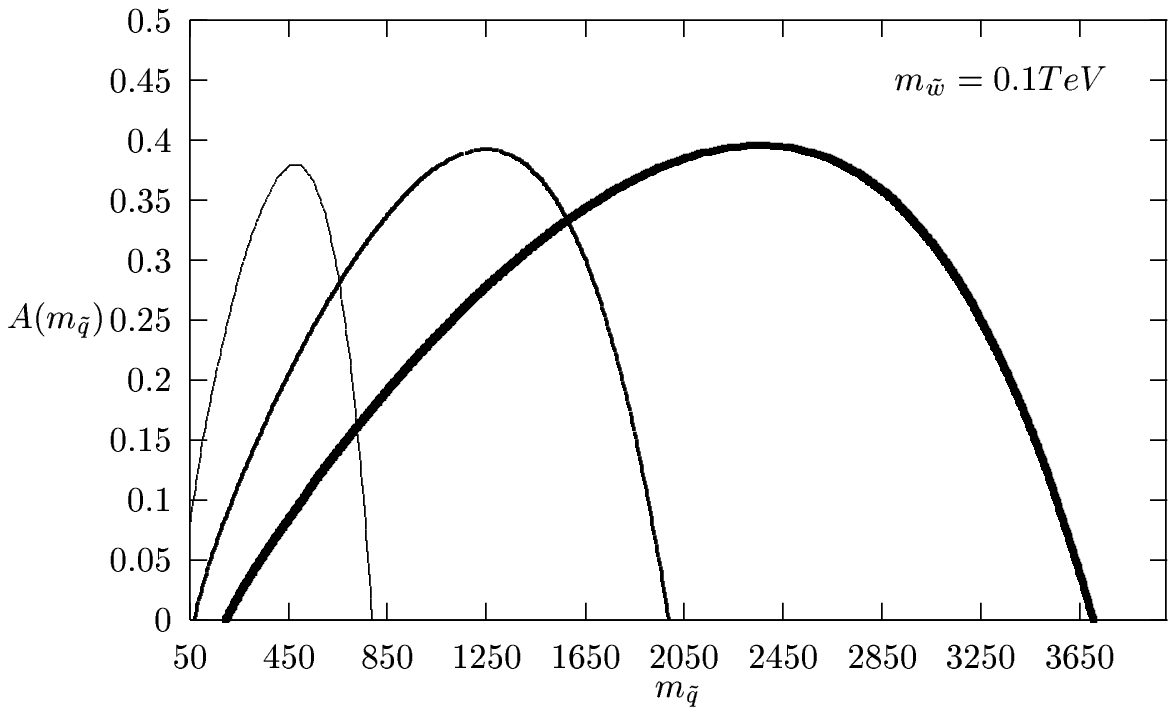}
\smallskip

Fig.8.c
\end{center}
\end{document}